\documentstyle[11pt,AATS]{article}

\begin{document}

\title{Dwarf spheroidals with tidal tails and structure formation}

\author{Lucio Mayer}
\affil{Department of Astronomy, University of Washington, Seattle
(USA)}

\begin{abstract}

Recent N-Body simulations have shown that dIrrs evolve into dSphs
owing to the strong tidal perturbation of the Milky Way.
Satellites whose dark matter halos have 
a core or an NFW profile with a concentration $c < 5$ 
undergo severe stripping even on low eccentricity orbits
and their remnants have projected stellar
profiles that become flat at large radii 
as observed in some of the dSphs. 
Satellites with higher central densities, as predicted by CDM models,
are more robust to tides and can reproduce 
the observations only on nearly radial orbits. Accurate 
proper motions for LG dSphs will allow to test current
models of structure formation.

\end{abstract}

\section{Introduction}

Dwarf galaxies in the Local Group show a striking morphology density 
relation; low surface brightness, gas poor dwarf spheroidals (dSphs),
that are supported by velocity dispersion of their stars,
are found within 300 kpc from either the MW or M31, while gas-rich,
rotationally supported dwarf irregulars (dIrrs) populate the 
outskirts of the LG  (Grebel 1999). 
Recently, high resolution N-Body simulations performed on supercomputers with
the parallel binary treecode PKDGRAV (Stadel \& Quinn, in preparation)
have shown that the strong, time-dependent tidal force exerted by the 
halo of the Milky Way induces severe mass loss and non-axisymmetric
instabilities in dIrrs, turning them into objects that match all
the observed properties of dSphs in less than 10 Gyr (Mayer et al. 2001a,b).
A signature left by ``tidal stirring'' are unbound 
stellar tails and streams that keep following the orbit
of the satellite for several Gyr.
Unbound stars lying or projected close to the
spheroidal remnant might be responsible for the outer flattening of the 
stellar surface density profiles of some dSphs, in particular Carina
(Irwin \& Hatzidimitriou  1995; Majewski et al. 2000). 
Here we will show that such features can be used to probe 
structure formation models.

\section{Initial Conditions}

We place N-Body models of dwarf irregular galaxies on bound orbits in 
the Milky Way halo, represented by the static external potential of an
isothermal sphere with a mass $M_{MW} \sim 4 \times 10^{12} M_{\odot}$,
a virial radius $R_{MW} \sim 400$ kpc and a core radius $R_c=4$ kpc
(the resulting circular velocity at the solar radius is $\sim 220$ km/s). 
The dwarf galaxy models comprise an exponential disk of stars embedded in
a dark matter halo. The sizes and masses of the halos and disks are
assigned according to both structure formation models and observations
(see Mayer et al. 2001b). 
The halo of the satellite  is either a truncated  isothermal sphere or
an NFW model with a concentration $c < 10$ 
(Navarro, Frenk \& White 1997).
The disk scale length $r_h$ is determined by the concentration (or the core
radius $r_c$ in the case of an isothermal halo) and  by the 
the spin parameter $\lambda$ of the halo (see Mo et al. 1998).
We consider different disk/orbit orientations and orbital eccentricities.

\begin{figure}[hb]	
\plotfiddle{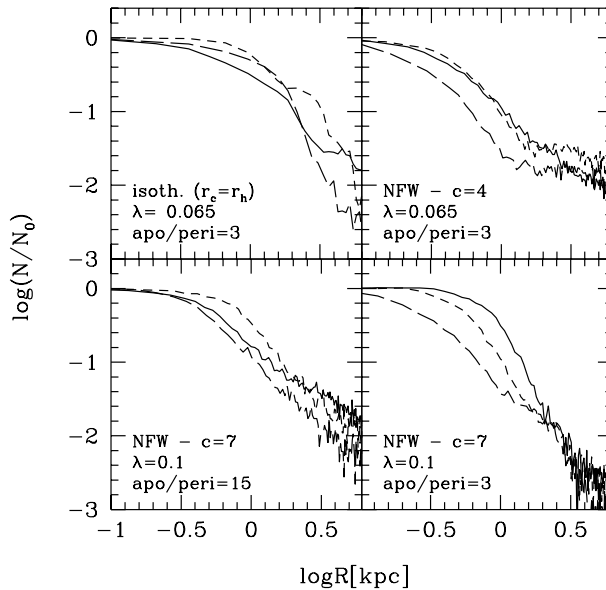}{7cm}{0}{45}{45}{-100}{-80}
\caption{Projected star counts for the remnants after 8 Gyr
for a line of sight
along one of the tidal tails (solid line) and along two directions
perpendicular to the latter (short and long dashed lines).
The orbital eccentricity and the models used are
also indicated (the apocenters fall between 120 and 280 kpc; as
a reference the present distance of Carina from the MW is $\sim 110$ kpc).
The initial mass of the satellites is $M_{sat}=10^{-4} M_{MW}$, 
and the angle between the spin of the disk and the orbital angular 
momentum is $\sim 63$ degrees.}
\end{figure}

\section{Results}

The remnants in our N-Body simulations exhibit projected profiles
with an outer flattening, like some of the dSphs, for a variety
of initial conditions; both the internal
structure and the orbits of the satellites determine 
the fraction of the stellar mass that is stripped and that can 
more or less alter the observed shape of the profile depending 
on the viewing angle (Figure 1).
However, when satellites have a shallow inner  density profile,
like that of a truncated isothermal sphere or that of
an NFW profile
with a low concentration ($c=4$), the observations can be reproduced
with orbits having a moderate eccentricity (apo/peri=3)
, whereas if they have NFW profiles with higher concentrations 
($c \ge 7$), as predicted by CDM models on their mass scale, 
nearly radial orbits are needed for a strong
mass loss to occur and lead to the flattening (Figure 1). 
This result holds even when the disk scale length is considerably
increased owing to a large halo spin parameter; the high concentration
moves the tidal radius of the system too further out unless the pericenter
of the orbit is very small.

\section{Conclusion}

So far, the few proper motion measurements available for LG dwarf galaxies
seem to indicate that the latter have low eccentricity orbits (e.g. the
Magellanic Clouds in Kroupa \& Bastian 1997 and Sculptor in Schweitzer
et al. 1995). If this result will be confirmed by the forthcoming
generation of high-precision astrometric missions, like GAIA and SIM,
the outer flattening observed in the stellar profiles of dSphs
will be hardly explained within CDM models. Recently, Eke et al. 
(2000) have shown that in warm dark matter models 
small galaxy halos have NFW profiles with $c < 5$;
these cosmologies would easily match the observations and, in addition, could 
yield the correct number of satellites in galaxy halos (Moore et
al. 1999) as well as halo profiles soft enough to reproduce 
the rotation curves of LSB galaxies (de Blok et al. 2000).

\acknowledgments

I thank Tom Quinn and Joachim Stadel for providing me with PKDGRAV
and Volker Springel for allowing me to use the code that builds
the galaxy models. I thank also Ben Moore, Fabio Governato and 
Monica Colpi for many stimulating discussions.
Simulations were carried out at CINECA and ARSC supercomputing centers.


\begin{references}


\reference
de Blok, W.J.G., McGaugh, S.S., Bosma, A., \& Rubin. V.C. 2001, astro-ph/0103102, submitted to ApJ


\reference
Eke, V.R., Navarro, J.F. \& Steinmetz, M., astro-ph/0012337, submitted to
ApJ

\reference
Grebel, E.K. 1999
IAU Symp. 192, Cape Town, in press, eds. Whitelock, P. \& Cannon

\reference
Irwin, M., \& Hatzdimitriou, D. 1995, MNRAS, 277, 1354

\reference
Kroupa P. \& Bastian U. 1997. New Ast 2, 139


\reference
Majewski, S.R., Ostheimer, J.C., Patterson, R.J., Kunkel, W.E., 
Johnston, K.V., \& Geisler, D. 2000, AJ, 119, 760

\reference
Mayer, L., Governato, F., Colpi, M., Moore, B., Quinn, T., Wadsley, J.,
Stadel, J., \& Lake, G. 2001a, ApJL, 547, L123

\reference
Mayer, L., Governato, F., Colpi, M., Moore, B., Quinn, T., Wadsley, J.,
Stadel, J., \& Lake G. 2001b, astro-ph/0103430, submitted to ApJ

\reference
Mo, H.J., Mao, S., \& White, S.D.M.  1998, MNRAS, 296, 847

\reference
Moore, B., Ghigna, S., Governato, F., Lake, G., Quinn, T.,
Stadel, J., \& Tozzi, P. 1999, ApJL, 524, L19-L22

\reference
Navarro, J.F, Frenk, C.S., \& White. S.D.M. 1997, MNRAS, 490, 493

\reference
Schweitzer AE, Cudworth KM, Majewski SR, Suntzeff NB. 1995. AJ 110, 2747

\end{references}
\end{document}